\documentclass{emulateapj}                                            %

\newcommand{\be}{\begin{equation}}
\newcommand{\ee}{\end{equation}}
\newcommand{\ba}{\begin{eqnarray}}
\newcommand{\ea}{\end{eqnarray}}

\newcommand{\saxj}{\mbox{SAX~J1808.4-3658 }}
\newcommand{\herx}{\mbox{Her~X-1 }}

\newcommand{\bigo}{\mathcal{O}}
\newcommand{\xhat}{{\bf \; x } }
\newcommand{\yhat}{{\bf \; y } }
\newcommand{\zhat}{{\bf \; z } }
\newcommand{\nhat}{{\bf  n } }
\newcommand{\rhat}{{\bf  r } }
\newcommand{\lvec}{{\bf l}}
\newcommand{\kvec}{{\bf k}}
\newcommand{\rbar}{\bar{r}}

\newcommand{\tilx}{\tilde{\zeta}}
\newcommand{\tily}{\tilde{\epsilon}}
 
\slugcomment{Accepted by the Astrophysical Journal}
\shorttitle{The Oblate Schwarzschild Approximation}

\begin{document}

\title{The Oblate Schwarzschild Approximation for Light Curves of 
Rapidly Rotating Neutron Stars}
\author {Sharon M. Morsink\altaffilmark{1,2}, Denis A. Leahy\altaffilmark{3},
Coire Cadeau\altaffilmark{1}, \& John Braga\altaffilmark{1}}

\altaffiltext{1}{Department of Physics,
University of Alberta, Edmonton AB,  T6G~2G7, Canada}
\altaffiltext{2}{On sabbatical at: Pacific Institute for Theoretical Physics,
Department of Physics and Astronomy, University of British  Columbia,
6224 Agricultural Road, Vancouver BC, V6T~1Z1, Canada}
\altaffiltext{3}{Department of Physics and Astronomy, University of Calgary,
Calgary AB, T2N~1N4, Canada}

\begin{abstract}
We present a simple method for including the oblateness of a rapidly rotating
neutron star when fitting X-ray light curves. In previous work we showed that
the oblateness induced by rotation at frequencies above 300 Hz produces 
a geometric effect which needs to be accounted for when modelling light curves
to extract constraints on the neutron star's mass and radius. In our model
 X-rays are emitted 
from the surface of an oblate neutron star and propagate to the observer
along geodesics of the Schwarzschild metric for a spherical neutron star.
Doppler effects due to rotation are added in the same manner as in the case 
of a spherical neutron star. We show that this model captures the most important
effects due to the neutron star's rotation. We also explain how the
geometric oblateness effect can rival the Doppler effect for some
emission geometries.   
\end{abstract}


\section{Introduction}
\label{s:intro}

The observation of pulsed X-ray emission originating from the surfaces
of rotating neutron stars provides an excellent opportunity to study the 
properties of neutron stars and to possibly constrain the equation of
state (EOS) of supernuclear density matter. Detailed modelling 
of the properties of the emitting region on the star combined with 
relativistic raytracing can be used to create theoretical light curves
which can then be compared with the observed light curves to
constrain the masses
and radii (and hence the EOS) of the X-ray emitting neutron stars. 

The best candidates for a pulse-shape analysis are the 
accreting neutron stars where the energy released from accretion
leads to X-ray emission from the surfaces of the neutron stars.
The accreting neutron stars that are of interest fall into
three classes: slowly rotating X-ray pulsars such as \herx;
accreting ms-Period neutron stars exhibiting Type I X-ray bursts;
and the accreting ms-Period X-ray pulsars such as \saxj.
(These classes are not mutually exclusive: some of the accreting
ms pulsars are also Type I bursters.)
In the case of \herx (with a spin period of 1.24 s), 
\citet{lea04} has shown that an accretion-column
model can explain the data and constrain the dimensionless
mass-to-radius ratio to the range $0.18 \le M/R\le 0.19$.
\citet{Bhat05} modelled the burst oscillations detected during
Type I X-ray bursts originating on XTE J1814-338 (spin period 3.2 ms) 
and found an upper limit of $M/R\le 0.24$. 
An analysis of the X-ray pulse shape of \saxj (spin period 2.5 ms)
by \citet{PG03} also provided constraints on the mass-to-radius
ratios to the range $0.18 \le M/R \le 0.3$. In addition, one non-accreting ms X-ray pulsar 
PSR J0437-4715 has been modelled by \citet{Bog07} and has
been shown to be consistent with a radius $R>6.7$ km. 
While none of these constraints are strong enough to conclusively rule out 
any equations of state, these studies do show a great deal of
promise.

The most commonly used approximation used to model neutron star light curves
is the Schwarzschild + Doppler (S+D) approximation \citep{ML98,PG03}. In the 
S+D approximation the gravitational light-bending effects are modelled as
though the star is not rotating using the Schwarzschild metric and the
formalism prescribed by \citet{PFC83}. Rotational effects are added by 
introducing Doppler terms as though the star is a rotating object with 
no gravitational field. 
In our previous 
calculations \citep{CMLC07} we found that as long as the star's spin
frequency is below 300 Hz, the ratio $M/R$ can be accurately
extracted using the S+D approximation.

In two recent papers \citep{CLM05,CMLC07} we
investigated the validity of the S+D approximation by computing 
the geodesics of rapidly rotating neutron stars using the exact numerical
metric to describe the gravitational field and the shape of the star. 
We have shown \citep{CLM05} that it is necessary to correctly bin the
light curve so that the variable times of arrival of photons from different parts
of the star are taken into account if the star is rapidly rotating. However,
we also showed that it was sufficient to use the Schwarzschild geometry to 
calculate the times of arrival: the effect of frame-dragging on light-bending
and times of arrival is not large enough to affect the light curves,
consistent with the calculations done by \citet{Braje00}.

In the second paper \citep{CMLC07} we investigated the effect of the neutron
star's oblate shape. Previous calculations \citep{Braje01} of
magnetospheric scattering found that the oblate shape of a rotating
neutron star can alter the directions that photons are scattered
into. In our computations (without scattering) we found that the oblate shape
has a significant effect on the light curves which should be included 
when modelling the light curves of rapidly rotating neutron stars. The 
reason for the importance of oblateness is simply geometrical: when light
is emitted from an oblate surface the directions that the light can be 
emitted into are different from those possible from the surface of 
a spherical star. This leads to certain spot locations on the star where
the spot is visible if the surface is oblate but would be invisible 
if the surface were spherical (and vice versa).

In our previous work \citep{CMLC07} we found that as long as the correct
shape of the star is used to formulate the initial conditions, the
light curves resulting from tracing rays in the Schwarzschild 
metric are a very good approximation to the light curves resulting
from ray-tracing in the full numerical metric of a rapidly
rotating neutron star. The problem with the method used in our past
work is that it is too slow for use in fitting light curves to 
theoretical models. In our past work a full relativistic stellar
structure calculation as well as geodesics connecting all spot
and observer latitudes had
to be computed for each hypothetical neutron star model. Since
many thousands of models must be computed in order to 
fit the neutron star's properties to a light curve, our
past work was not useful for comparing theoretical calculations
with X-ray timing data. 

The purpose of this paper is to present a simple approximation
that includes the essential features of our more detailed 
ray-tracing calculations
for rapidly rotating neutron stars. Our new approximation,
the Oblate Schwarzschild (OS) approximation, is based 
on the more commonly used S+D approximation. In the OS 
approximation an empirical formula for the oblate shape of
the rotating neutron star is used to define initial conditions 
and visibility conditions for photons. Once initial conditions
on the surface of the star are set up, the Schwarzschild 
metric is used to find bending angles and photon time delays. 
Doppler effects due to rotation are then added to the model
in the same manner as in the S+D approximation. 

The outline of this paper is as follows. In Section~\ref{s:oblate}
we present a simple empirical formula for the shape of the star
and derive expressions for the vector normal to the surface, angles
between the surface normal and the light rays, and the solid
angle subtended by an oblate surface. In Section~\ref{s:light}
we explain how these concepts are used to create light curves in
the OS approximation, providing details of how the new visibility
condition is implemented. In Section~\ref{s:effect} we explain
why the geometric effect of oblateness can rival the Doppler
effect for some emission geometries.
 In Section~\ref{s:compare} we compare
our approximate light curves with light curves computed using our
previous exact methods. Finally, we conclude with comments on
possible applications of the OS approximation.

\section{The Oblate Shape of a Rotating Neutron Star}
\label{s:oblate}

In the OS approximation, an oblate surface is embedded in the metric of
a non-rotating neutron star. The metric exterior to a non-rotating neutron star
is given by the Schwarzschild solution
\begin{eqnarray}
ds^2 &=& - \left(1-\frac{2M}{r}\right) dt^2 +
\left(1-\frac{2M}{r}\right)^{-1} dr^2 \nonumber \\
&& + r^2 \left( d\theta^2 + \sin^2\theta d\phi^2
\right).
\end{eqnarray}
(We use gravitational units where $G=c=1$.)
The surface of the star is described by a function $r = R(\theta)$, and will be specified in 
subsection \ref{s:model}. Given a function describing the surface of the star, 
the surface area
of a small spot
of angular extent $d\theta$ and $d\phi$ located at
angles $\theta$ and $\phi$ on the surface of the star will have a surface 
area
\be
dS(\theta) = R^2(\theta) \sin\theta \left( 1 + f^2(\theta) \right)^{1/2} d\theta d\phi
\label{eq:dS}
\ee
where the function $f(\theta)$ is defined by
\be
f(\theta) = \frac{(1-2M/R)^{-1/2}}{R} \frac{dR}{d\theta} = \frac{(1+z)}{R}\frac{dR}{d\theta},
\label{eq:f}
\ee
where the gravitational redshift $z=1/\sqrt{1-2M/R} - 1$ has been introduced.

An alternative (but equivalent) description of the gravitational field exterior to 
a non-rotating star is given by the Isotropic Schwarzschild metric
(see, for example, \cite{MTW})
\begin{eqnarray}
ds^2 &=& - \left(\frac{(1-M/2\rbar)}{(1+M/2\rbar)}\right)^2 dt^2 
\nonumber \\
&&+ 
\left(1+M/2\rbar\right)^4 \left(d\rbar^2 + \rbar^2(d\theta^2 + \sin^2\theta d\phi^2)\right),
\label{eq:iso}
\end{eqnarray}
where the isotropic radial coordinate $\rbar$ is related to the areal radial coordinate $r$
by the transformations
\be
r = \rbar\left(1+M/2\rbar\right)^2 
\label{eq:rbar}
\ee
and 
\be
dr = \left(1-2M/r\right)^{1/2} (1+M/2\rbar)^2 d\rbar.
\label{eq:dr}
\ee
The surface 
of the star in this coordinate system is denoted $\rbar=\bar{R}(\theta)$.
Making use of the coordinate transformations given by equations (\ref{eq:rbar}) 
and (\ref{eq:dr})
it is easily shown that the function $f(\theta)$ defined in equation (\ref{eq:f}) is
\be
f(\theta) = \frac{1}{\bar{R}} \frac{d\bar{R}}{d\theta}.
\label{eq:f2}
\ee

\subsection{A Model for Oblateness}
\label{s:model}
In order to find the shape of a rotating neutron star, it is necessary to specify an
equation of state (EOS) of dense nuclear matter,  a spin period and
mass and then solve the relativistic equations of stellar structure. The relativistic
stellar structure equations must be solved numerically using an axisymmetric
code such as {\tt rns}\footnote{Available at {\tt http://www.gravity.phys.uwm.edu/rns}}
 \citep{SF95}, based on the methods described by
\citet{CST94}. Once the stellar structure equations have been solved, the location
of the star's surface is extracted by finding the locations where the fluid's 
enthalpy vanishes. 

In order to 
aid the process of fitting light curves, we have found a simple empirical formula 
for the shapes of rotating neutron stars. We assume that the surface of a rotating
neutron star can be described by a function of the form
\be
\frac{R(\theta)}{R_{eq}} =  1 + \sum_{n=0}^{2} a_{2n}(\zeta,\epsilon) P_{2n}(\cos\theta)
\label{eq:shape1}
\ee
where
$\theta$ is the co-latitude, measured from the spin axis ($\cos\theta = 0 $ on the equator),
$P_n(\cos\theta)$ is the Legendre polynomial of order $n$, 
$R_{eq}$ is the radius of the rotating star measured at the equator and the parameters 
$\zeta$ and $\epsilon$ 
are given by
\be
\zeta = \frac{GM}{R_{eq}c^2}
\label{eq:zeta}
\ee
and 
\be
\epsilon = \frac{\Omega^2 R_{eq}^3}{GM} = \frac{\Omega^2 R_{eq}^2}{c^2} \frac{1}{\zeta}
\label{eq:epsilon}
\ee
with $\Omega = 2\pi/P$, where $P$ is the spin period. 
The coefficients $a_n(\zeta,\epsilon)$ can then be fit to the shapes of 
stellar models computed 
using the {\tt rns} code. The \citet{Hartle67} slow rotation approximation
corresponds to dropping the $a_4$ term in the series. We find that the $a_4$ term
is required to describe the shape of the largest, most rapidly rotating stars. However,
higher order terms are not required. 

We have computed the stellar structure of compact stars for a wide variety of equations of
state. For each EOS, models spanning the allowed values of mass and angular velocity
were computed. We find that the EOS can be grouped into two families, where each family's
stars can be described by a separate set of Legendre coefficients.  
The first family, Neutron and Hybrid Quark Stars, includes the following EOS: 
\citet{AB77} catalogue EOS A, B, C, F, G, L, N, O;
EOS APR \citep{APR98} which includes nuclei scattering data and special relativistic 
corrections; 
hybrid quark stars with a mixed quark-hadron phase calculated by \citet{ABPR};
and Hyperon stars computed by \citet{LNO06} using methods described by \citet{Glen}.
The second family consists of color-flavor-locked (CFL) quark stars described by
the EOS $P=(\epsilon-4B)/3$ where $B$ is the bag constant. Bag
constants in the range $50 - 140$ MeV$/\hbox{fm}^3$ are used to generate different 
CFL EOS. This simple EOS
has been shown by \citet{Fraga} to be an excellent approximation to quark star models
including second order perturbative QCD corrections.

Although each EOS has a different mass versus radius curves, for any given combination
of mass, radius and angular velocity the scaled shape of the star is independent of the
EOS within its family. The values of the Legendre 
fitting coefficients $a_n(\zeta,\epsilon)$ for the two families are given in Table 1.
Since $R(\pi/2)$ is defined to be the equatorial radius of the rotating star,
this requires that the coefficients obey $0=a_0-a_2/2+3a_4/8$. As can be seen from
the coefficients in Table 1 this requirement is not exactly satisfied.
However, the errors in this model are still less than 1\%.

Equation (\ref{eq:shape1}) is only useful if the equatorial radius of the star is known. 
When doing light curve fits, we will need to choose a value of $\theta$ describing 
the spot's co-latitude and the value of radius at the spot's latitude. Given these
values, equation (\ref{eq:shape1}) can be inverted to solve for $R_{eq}$,
\begin{eqnarray}
\frac{R_{eq}}{R(\theta)} &=&   1 + \sum_{n=0}^2 b_{2n}(\tilde{\zeta},\tilde{\epsilon}) 
P_{2n}(\cos\theta) \nonumber \\
&&  + P_{2}(\cos\theta) \sum_{n=1}^2    c_{2n}(\tilde{\epsilon}) P_{2n}(\cos\theta) ,
\label{eq:inverseshape}
\end{eqnarray}
where $\tilde{\zeta} = GM/R(\theta)c^2$ and $\tilde{\epsilon} = \Omega^2 R^3(\theta)/GM$.
The values of the $b_n$ and $c_n$ coefficients are displayed in Table~1. 
It is our intention that equation (\ref{eq:inverseshape})  be
used in light curve fitting to put constraints on the equatorial
radius of a rotating star. In order to compare with the predictions of
a specific EOS, it is still necessary to construct a sequence of
neutron stars with a specific spin frequency. The construction of
such a sequence for a fixed spin frequency can be done in a
straight-forward fashion using a public domain program such as
{\tt rns}.

\subsection{The Normal Vector}

The isotropic form of the Schwarzschild metric given in equation 
(\ref{eq:iso}) has the useful feature that surfaces of
constant time are conformally flat. This allows the introduction of quasi-Cartesian
coordinates $\{x,y,z\}$ defined by $x=\rbar\sin\theta\cos\phi$, $y=\rbar\sin\theta\sin\phi$
and $z=\rbar\cos\theta$. In the quasi-Cartesian coordinate system
the $z$ axis is aligned with the star's spin axis, and the
observer lies in the x-z plane at an angle $i$ from the spin axis.

The vector normal to the surface at any arbitrary point can be constructed using the 
quasi-Cartesian coordinate system in the same way that it would be computed in flat space. 
First a tangent vector is constructed by taking the gradient of the function $\bar{R}$
making use of equation (\ref{eq:f2}). 
The normal to the surface at the location of the spot is 
\begin{eqnarray}
\nhat &=& \left( 1 + f^2(\theta) \right)^{-1/2}
\left[
\left(\sin\theta - \cos\theta f(\theta) \right) ( \cos\phi \xhat +
\sin\phi \yhat ) \right.
\nonumber \\
&&\left. + \left(\cos\theta +  \sin\theta f(\theta) \right) \zhat \right].
\label{eq:nhat1}
\end{eqnarray}
The unit radial vector is defined by $\rhat = \sin \theta (\cos\phi
\xhat + \sin\phi\yhat) 
+ \cos \theta \zhat$,
so the angle $\gamma$ between $\rhat$ and $\nhat$,
is simply
\be
\cos \gamma =  \left( 1 + f^2(\theta) \right)^{-1/2}.
\label{eq:cgamma}
\ee
Since $\sin \gamma = f/\sqrt{1+f^2}$, the normal vector in equation (\ref{eq:nhat1}) can also be expressed as
\be
\nhat = \sin(\theta-\gamma) \left( \cos\phi \xhat + \sin\phi \yhat\right)
+ \cos(\theta-\gamma) \zhat
\label{eq:nhat2}
\ee
as would be expected from simple geometric considerations in flat space.

\subsection{Bending Angles and Zenith Angles}
\label{s:bend}

A light ray emitted in an initial direction $\lvec$ will be moving a different 
direction $\kvec$ when it
reaches the observer. Since the observer is located at an angle $i$ from the spin axis,
$\kvec = \sin i \xhat + \cos i \zhat$. In the standard treatment of spherical 
neutron stars, $\alpha$ is the angle between the initial photon direction and the
radial direction, defined by $\cos\alpha = \lvec \cdot \rhat$
and the bending angle $\psi$ is defined by 
\be 
\cos \psi = \kvec \cdot \rhat = \cos i \cos \theta + \sin i \sin \theta \cos \phi.
\label{eq:cpsi}
\ee
In the Schwarzschild metric, the vectors $\lvec$ and $\kvec$ are 
coplanar so that 
\be
\lvec = \frac{1}{\sin \psi} \left( \sin\alpha \kvec + \sin(\psi-\alpha) \rhat\right).
\label{eq:lvec}
\ee

\begin{figure}
\begin{center}
\plotone{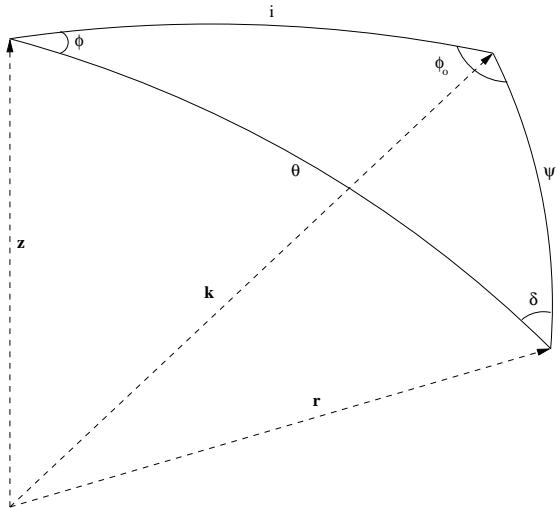}
\end{center}
\caption[]{Spherical triangle illustrating the emission geometry.
The vector $\zhat$ is aligned with the spin axis, $\rhat$ points
in the direction of the spot, and $\kvec$ points in the direction
of the observer. The angles between these three vectors as
well as the interior angles of the spherical triangle are shown.
}
\label{fig:trig}
\end{figure}

The zenith angle $\beta$ between the normal vector and the initial
photon direction is defined by $\cos\beta = \lvec \cdot \nhat$. Making
use of equation (\ref{eq:lvec}), the identity $\nhat =
(\sin(\theta-\gamma) \rhat + \sin\gamma\zhat)/\sin\theta$, and
the spherical trigonometric identity $\cos i =\cos\theta\cos\psi +
\sin\theta \sin\psi \cos\delta$ (see Figure \ref{fig:trig} for 
the geometrical definition of $\delta$)
the zenith
angle has the value
\be
\cos \beta =  \lvec \cdot \nhat = \cos\alpha \cos\gamma + \sin\alpha\sin\gamma \cos\delta,
\label{eq:cbeta}
\ee
if $\sin\psi \neq 0$. If $\sin\psi=0$, it is straight forward to show that
$\cos \beta = \cos\alpha \cos\gamma$.

\subsection{Solid Angle for Static Oblate Stars}

In this section we derive the solid angle subtended by an surface area element on an
oblate star that is at rest. In the subsequent section we will discuss the modifications
necessary when the star is rotating.

Since the photons travel to the observer via geodesics of the Schwarzschild metric,
the solid angle subtended by a surface element is given by the standard 
expression \citep{PFC83}
\be
d\Omega_s = \frac{b \;db \;d\phi_o}{D^2}
\ee
where $b$ is the photon's impact parameter, $D$ is the distance between the 
star and the observer,
and $\phi_o$ is the azimuthal angle around the vector $\kvec$. 
The subscript ``s'' refers to static, since in this section we 
do not take into account Doppler shifts and boosts due to the rotation of the star.
The impact 
parameter is related to the zenith angle $\alpha$ by
\be
\sin \alpha = \frac{b}{R(\theta)} \sqrt{1-\frac{2M}{R(\theta)}}.
\label{eq:sinalpha}
\ee
The bending
angle $\psi$ is related to the impact parameter and the surface of the star
by
\be
\psi(b,R(\theta)) = b \int_{R(\theta)}^\infty \frac{dr}{r^2} 
\left( 1 - \frac{b^2}{r^2} (1-\frac{2M}{r})\right)^{-1/2}.
\label{eq:psi}
\ee

In order to express the solid angle in terms of the coordinates defined 
on the surface of the star, a change of variables from coordinates $\{b, \phi_o\}$ 
to $\{\psi,\phi_o\}$ must be made so that 
\be
d \Omega_s = \left(\frac{\partial b}{\partial \psi}\right)_{\!\phi_o} \frac{b d\psi d\phi_o}{D^2}. 
\label{eq:dOm2}
\ee
The relationship between the impact parameter and the bending angle
for a spherical star is simple, since the photons are emitted from a
surface of constant gravitational potential. In the case of an oblate star,
the bending angle $\psi$ depends on both the impact parameter 
and the variable location of the star's surface. 
The partial derivative appearing in equation (\ref{eq:dOm2}) is
\be
\left(\frac{\partial b}{\partial \psi}\right)_{\!\phi_o} = \left(\frac{\partial b}{\partial \psi}\right)_{\!R}
+ \left(\frac{\partial b}{\partial R}\right)_{\!\psi} \left(\frac{\partial R}{\partial \psi}\right)_{\!\phi_o}
\label{eq:partial1}
\ee
where  $\left( {\partial b}/{\partial \psi} \right)_R = 1/\left( {\partial \psi}/{\partial b} \right)_R$
is evaluated using the integral definition given in equation~(\ref{eq:psi}).
Making use of standard relations for partial derivatives
\be
 \left(\frac{\partial b}{\partial R}\right)_{\!\psi} = 
- 
\left( \frac{\partial b}{\partial \psi} \right)_{\!R}
\left( \frac{\partial \psi}{\partial R} \right)_{\!b}.
\label{eq:partial}
\ee
Using equations (\ref{eq:sinalpha}) and (\ref{eq:psi}) the second partial derivative 
appearing on the right-hand side of 
equation (\ref{eq:partial}) is
\be
\left(\frac{\partial \psi}{\partial R}\right)_{\!b} = 
- \frac{(1+z)}{R} \frac{\sin\alpha}{\cos\alpha}.
\ee
The partial derivative of $R$ with respect to $\psi$ 
can be calculated using equations (\ref{eq:f}) and (\ref{eq:cgamma}),
\be
\frac{1}{R}  \left( \frac{\partial R}{\partial \psi} \right)_{\!\phi_o}    
= - \frac{f(\theta)}{(1+z)\sin \theta} 
\left(\frac{\partial \cos \theta}{\partial \psi}\right)_{\!\phi_o}.
\ee
Making use of standard identities for the spherical triangle defined
in Figure \ref{fig:trig}, 
\be
\left( \frac{\partial \cos\theta}{\partial \psi}\right)_{\!\phi_0}
= -\sin \theta \cos \delta.
\ee
The $\cos \delta$ term in this last expression can be eliminated
using equation (\ref{eq:cbeta}), leading to the simple equation
\be
\frac{1}{R}  \left( \frac{\partial R}{\partial \psi} \right)_{\!\phi_o}  
= \frac{ \left(\cos \beta - \cos\alpha \cos\gamma\right)}{(1+z)\cos\gamma \sin \alpha}.
\ee

The final result for equation (\ref{eq:partial1}) is
\be
\left(\frac{\partial b}{\partial \psi}\right)_{\!\phi_o} 
= \left(\frac{\partial b}{\partial \psi}\right)_{\!R} \cos \beta.
\ee

The expression for the solid angle can now be simplified to 
\be
d\Omega_s = (1+z)^2 \frac{R^2}{D^2} \cos\beta 
\left|\frac{\partial \cos\alpha}{\partial \cos\psi}\right|_{\!R}
\frac{\sin \psi}{\cos\gamma} d\psi d\phi_o.
\ee
The transformation between the $\{\psi, \phi_o\}$ coordinates to 
the star's coordinates $\{\theta,\phi\}$
is given by the usual relation $\sin\psi d\psi d\phi_o = \sin \theta d\theta d\phi$.  
Equation (\ref{eq:dS})
can now be used to simplify the solid angle formula to 
\be
d\Omega_s = (1+z)^2 \frac{dS}{D^2}  \cos\beta 
\left|\frac{\partial \cos\alpha}{\partial \cos\psi}\right|_R.
\label{eq:dOm}
\ee

The solid angle subtended by a spot on an oblate star given by 
equation (\ref{eq:dOm}) 
is very similar to the expression for the
solid angle subtended by a spot on a spherical surface. The 
expressions for the oblate and spherical surfaces differ through
the $\cos \beta$ term, corresponding to the cosine of the
angle between the initial photon direction and the vector
normal to the surface. In the spherical case, $\cos\beta$ is
replaced by $\cos\alpha$, the cosine of the angle between the
initial photon direction and the radial vector. Hence our
result for the solid angle reduces to a simple correction for the
tilt of the surface induced by oblateness.

\section{Light Curves for Rotating Oblate Stars}
\label{s:light}

The flux due to a spot on the star is 
\be
dF= I_o  d\Omega_o
\ee
where  $I_o$ is specific intensity measured
in the observer's frame and $d\Omega_o$ is the solid angle subtended by  the spot as
measured by the observer.

Light emitted with frequency $\nu_{em}$ will detected with frequency $\nu_o$ given
by
\be
\nu_o = \frac{\eta}{1+z} \nu_{em}
\ee
where $\eta$ is the Doppler boost factor given by
\be
\eta = \frac{\sqrt{1-v^2}}{1-{v}\cos \xi},
\label{eq:boost}
\ee
where $v$ is the magnitude of the spot's velocity 
and $\xi$ is the angle between the velocity and the original photon direction.
The spot's velocity is unaffected by the oblate shape of the star, so the 
functions $v$ and $\xi$ are given by the values
\be
v = 2 \pi \nu_* R(\theta) \sin \theta \left(1 - \frac{2M}{R}\right)^{-1/2}
\ee
and 
\be
\cos \xi = - \frac{\sin\alpha\sin i \sin \phi}{\sin\psi}.
\label{eq:cosxi}
\ee

The motion of the spot causes an aberration in the angles between the normal to the
surface and the initial photon direction as measured in the rest frame of the spot,
$\beta_{em}$ and the observer at infinity, $\beta$, given by
\be
\cos\beta_{em} = \eta \cos\beta
\ee
where $\beta$ is given by equation (\ref{eq:cbeta}). In order for the spot to 
be visible, it is necessary that $0\le\cos\beta_{em}\le 1$. 
The solid angle measured at infinity $d\Omega_o$ is also affected by the 
spot's motion, and is related to the solid angle derived in the 
previous section by 
\be
d\Omega_o = \eta d\Omega_s.
\ee

A photon emitted from
$R(\theta)$ and with impact parameter $b$ will arrive at a time
\begin{eqnarray}
T(b,R(\theta)) &=& \int_{R(\theta)}^\infty dr
\left(1-\frac{2M}{r}\right)^{-1}
\times \nonumber \\
&&\left( \left(1 - \frac{b^2}{r^2}\left(1-\frac{2M}{r}\right) \right)^{-1/2}-1\right) 
\label{eq:toa}
\end{eqnarray}
relative to a photon with zero impact parameter emitted from the same location. 

Finally, making using of the relativistic conservation law $I_o/\nu_o^3 = I/\nu_{em}^3$,
the measured flux  is 
\be
F_{\nu_o} = \int_{\cos\beta_{em}>0} d\Omega_s \eta^4 (1+z)^{-3} I(\cos\beta_{em},\nu_{em})
\ee
where $d\Omega_s$ is given by equation (\ref{eq:dOm}). In order to produce a light curve
the flux should be integrated over the observed frequency range, and the flux should
be binned according to the relative arrival times of the
photons. Times of arrival can be computed exactly using equation
(\ref{eq:toa}) or a series expansion such as that given by \cite{PB06}
(see their equation 18) can be used. 

\subsection{Initially Ingoing Photons}

\begin{figure}
\begin{center}
\plotone{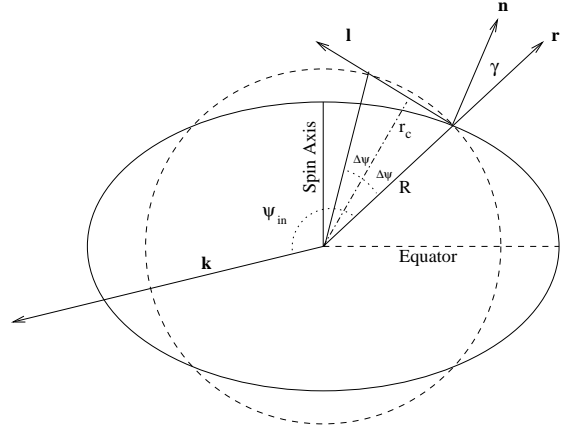}
\end{center}
\caption[]{Side view of a rotating neutron star (solid curve).
The surface of a spherical star with the same radius as the oblate star
at the emission latitude is shown as a dashed circle. 
A photon with initial direction $\lvec$ is initially moving towards
smaller values of the radial coordinate. At $r_c$ the photon reaches
the minimum value of $r$ and begins to move outwards. The change in
bending angle while the photon travels between the two intersections of the
solid and dashed curves is $2\Delta \psi$.
}
\label{fig:star}
\end{figure}

For photons that are emitted in an outgoing direction, the calculation of the flux is
straightforward and similar to the standard calculation for a star with a spherical
surface. However, because the surface is oblate, there will be photons emitted which
are initially directed radially inwards. These photons will travel to smaller values 
of $r$ until a critical radial coordinate $r_c$ is reached, at which point $\dot{r}=0$. 
After the critical radius has been reached, the photon will move outwards. In
Figure~\ref{fig:star} an initially ingoing photon's trajectory is labelled 
by vector $\lvec$.

An initially ingoing photon will have a value of impact parameter 
\be
b_{in} \le \frac{R}{\sqrt{1-2M/R}},
\ee
and an angle $\alpha > \pi/2$. The quantity $R$ is the location of the surface of the
star at the latitude of the spot.
 The critical radius is given by the solution of the
equation
\be
r_c =  b_{in} \sqrt{1-\frac{2M}{r_c}}.
\ee
Given this critical radius, the bending angle associated with the motion of
the photon from $R$ to $r_c$ is $\Delta\psi$ given by the integral
\be
\Delta\psi = b_{in} \int_{r_c}^{R} \frac{dr}{r^2} 
\left( 1 - \frac{b_{in}^2}{r^2} (1-\frac{2M}{r})\right)^{-1/2}.
\ee
Similarly, the relative coordinate time taken by the photon to travel this
distance is
\be
\Delta T =  \int_{r_c}^R dr \left(1-\frac{2M}{r}\right)^{-1}
\left(1 - \frac{b_{in}^2}{r^2}\left(1-\frac{2M}{r}\right) \right)^{-1/2} 
\ee
By symmetry, when the photon travels from $r_c$ out to $R$ the bending
angle is $\Delta \psi$
and the extra time is $\Delta T$. The final result is that the total bending angle for an
initially ingoing photon is
\be
\psi_{in}(b_{in},R) = 2 \Delta \psi + \psi(b_{in},R)
\ee
where $\psi(b,R)$ is given by equation (\ref{eq:psi}). Similarly, the
relative arrival time for the photon is
\be
T_{in}(b_{in},R) = 2 \Delta T + T(b_{in},R).
\ee
As expected, an initially ingoing photon will have a larger bending angle and take
 longer to reach the
observer than an initially outgoing photon with the same value of impact parameter. 

In all situations of interest, the difference $R(\theta)-r_c$ will always be much 
smaller than the 
radius of the star. This makes it possible to approximate the integrals 
$\Delta \psi$ and $\Delta T$ by 
assuming that in the integrand $r=r_c(1+\varepsilon)$ where 
$\varepsilon \ll 1$. 
To lowest order in
$\varepsilon^{1/2}$, 
\be
\Delta \psi = \sqrt{2\frac{(R-r_c)}{(r_c-3M)}}.
\ee
Similarly an approximation for $\Delta T$ is
\be
\Delta T = {r_c} (1-2M/r_c)^{-1/2} \Delta \psi.
\ee

\section{Effect of Oblateness on Light Curves}
\label{s:effect}

It may seem counter-intuitive that the oblateness of a rotating star 
could have an important effect on the resulting light curves since a naive
argument predicts that oblateness is unimportant: since oblateness
is an effect that is second-order in angular velocity it ought to be
smaller than first-order effects such as Doppler effects and the 
frame-dragging effect. However, \citet{CMLC07} have shown that 
the effect of oblateness
can be large in some cases. The main reason why oblateness 
can't be neglected is {\em not} through its effect on the 
star's gravitational field. It is instead a simpler 
geometric argument that would also hold in flat space.
We now explain why the naive
argument fails for certain ranges of spot locations.

Doppler shifts and boosts are first-order corrections
to the shape of a light curve, so Doppler corrections are the most
important effect. As has been shown elsewhere (for example \citet{PG03})
Doppler boosting creates an asymmetry in the light curve. However,
Doppler effects are most important when the spot is moving in the 
same direction as the initial photon direction. For light emitted near the
back of the star (for values of rotational phase near $\phi=\pi$) the
angle $\xi$ defined in equation (\ref{eq:cosxi}) is close to $\pi/2$ 
so that the boost factor in equation (\ref{eq:boost}) only differs from
unity by corrections that are second order in $v/c$. It is in this region
near the back of the star that oblateness effects become most important.

Frame-dragging is almost always a more significant contribution to the
gravitational field than the contributions to the field due to
oblateness. Only at very high spin frequencies when $\epsilon \sim
0.1$ (corresponding to about 500 Hz, depending on the EOS) 
do the oblateness corrections to
the spacetime metric become comparable to the frame-dragging
corrections. However, neither correction has a strong influence on the
trajectories of photons or the resulting light curves. 
As the photon moves away from the star,
the frame-dragging frequency falls off as $r^{-3}$ so it rapidly
becomes unimportant.
(Corrections due to oblateness fall off even faster.)
 In our previous
calculations \citep{CMLC07} we found very little difference between
the light curves computed with the Kerr or Scwarzschild metric.
For this reason, the use of the Kerr metric (which keeps the frame-dragging term
in the metric) is not a great improvement over the Schwarzschild metric (which
ignores frame-dragging) when computing light curves.

Oblateness affects lights curves in two ways: the solid angle is changed since
the emitting surface is tilted at a different angle with respect to the
initial photon direction; and the visibility condition is altered. The first
effect is usually small (except near the back of the star) 
while the second effect can be surprisingly large.

The first oblateness effect is easily seen in equation (\ref{eq:dOm}) for 
the solid angle of an oblate star. This expression is equivalent to the
expression for the solid angle of a spherical star multiplied by the factor
$\cos\beta/\cos\alpha$. This factor corrects for the change in the tilt of
the surface due to oblateness. Since $\cos\beta$ (the 
angle between the surface normal and the initial photon)
 only differs from
$\cos\alpha$ (the angle between the radial direction and the initial photon)
by an amount of order $\bigo(\Omega^2)$ this correction is not
typically large. However, near the back of the star (in relation to
the observer) this correction factor can be important if $\cos\alpha$
is of similar order of magnitude as $\gamma$ or smaller. In order to
see this, consider the value of phase $\phi=\pi$, corresponding to the
back of the star. At this point, equation (\ref{eq:cbeta}) for the zenith angle
reduces to $\cos\beta = \cos\alpha \cos\gamma + \sin\alpha
\sin\gamma$. If $\cos\alpha$ is small and $\bigo(\cos\alpha) \sim
\bigo(\gamma)$ then $\cos\beta/\cos\alpha \sim 1 + \gamma/\cos\alpha +
\bigo(\gamma^2)$. In this case the correction factor
$\cos\beta/\cos\alpha$ is not necessarily close to unity, and can 
cause large changes in the received flux. Similar arguments hold for
other values of phase as long as $\cos\alpha$ is small.

The second effect of oblateness, due to the change in the visibility
condition, can be understood with a simple example. Consider a spot
at fixed latitude $\theta$ on two equal-mass stars, a spherical
star and an oblate star, chosen so that the radii of the two stars at the
spots' latitude are the same. Suppose that at the value of rotational phase denoted
$\phi_1$ the spot is at the limb of the spherical star. This means that $\alpha=\pi/2$
and the bending angle is at its maximum value, $\psi_{max}$. Using equation 
(\ref{eq:cpsi}) $\cos\psi_{max}=\cos i \cos\theta + \sin i \sin\theta\cos\phi_1$.

Similarly, on the oblate star, the spot is at the limb at rotational phase 
$\phi_2=\phi_1+\Delta \phi$
at which point $\beta = \pi/2$ and the angle between the radial direction and
the initial photon direction is $\alpha_2$, differing by $\pi/2$ by an amount
second order in angular velocity. The bending angle for light emitted from
$\phi_2$ is $\cos\psi_2 = \cos i \cos\theta + \sin i \sin\theta \cos\phi_2$. 
The difference between the rotational phases on the two stars where the
spot is at the limb is given by 
\be
\cos\phi_2- \cos\phi_1 = \frac{\cos \psi_2 - \cos\psi_{max}}{\sin i \sin \theta}
= \left(\frac{d\cos\psi}{d\cos\alpha}\right)_{\!\alpha_2}
\frac{\cos\alpha_2}{\sin i \sin\theta}.
\label{eq:dcosphi}
\ee
If we assume that the phase difference, $\Delta \phi$ is very small, and that 
the spot is very close to the back of the star so that $\phi_1 \sim \pi$, then
$\cos\phi_2- \cos\phi_1 = (\Delta \phi)^2/2$. 
As a result, the change of phase over which the spot is visible due to 
oblateness is approximately
\be
\Delta \phi \sim \left( \frac{2}{\sin i \sin \theta} 
\left(\frac{d\cos\psi}{d\cos\alpha}\right)_{\!\alpha_2} \cos\alpha_2 \right)^{1/2}
\ee
when the spot is near the back of the star. In the \citet{Bel02} 
approximation $d\cos\psi/d\cos\alpha \sim (1+z)^2$ for light emitted
close to the normal to the surface. In our example light is emitted 
close to the tangent to the surface, but the cosine derivative 
continues to be finite and order unity. Since $\cos\alpha_2 \sim \bigo(\Omega^2)$
it follows that the phase difference $\Delta \phi \sim \bigo(\Omega)$. 
As a result, for light emitted on the side of the star opposite the observer,
the phase differences caused by oblateness are first order in the angular 
velocity. Since the Doppler boosting is only second order in the angular 
velocity in this region of the star, oblateness becomes the dominant effect.

\section{Comparison of Models}
\label{s:compare}

The effect of oblateness is most important for very large neutron
stars rotating with very high spin frequencies. In order to test the
accuracy of the OS approximation discussed in this paper, we now 
examine the results of this approximation for one very oblate 
neutron star with a mass of $1.4 M_\odot$, an equatorial
radius of $R_{eq}=16.4$ km and a spin frequency of 600~Hz. 
The expansion parameters defined in equations (\ref{eq:zeta})
and (\ref{eq:epsilon}) take on values of $\zeta=0.126$ and
$\epsilon=0.339$ for this model.

\begin{figure}
\begin{center}
\plotone{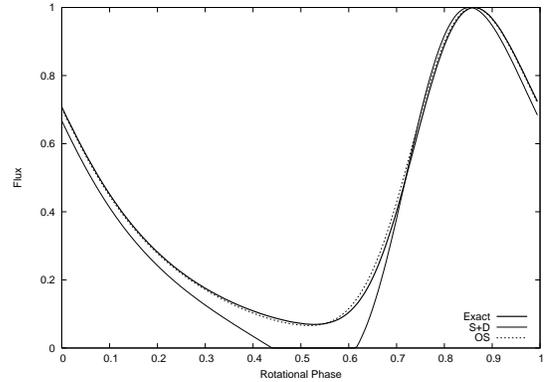}
\end{center}
\caption[]{Light curves for light emitted from a co-latitude of
  $\theta=49^\circ$ and an observer at an inclination of $i=70^\circ$.
The neutron star has $M=1.4 M_\odot$,  $R_{eq}=16.4$ km and spins at 600~Hz. 
The solid bold curve is computed using the exact numerical space time,
as described by \cite{CMLC07}. The solid curve labelled S+D is
  computed by assuming that the star's surface is spherical. The
  dashed curve is computed using the Oblate Schwarzschild
  approximation described in this paper. 
}
\label{fig:L49}
\end{figure}

Consider first the case of emission from a spot at co-latitude
$\theta=49^\circ$. The radius of the star (computed using
the exact numerical method) at this point is
$R(\theta)=15.1$ km. The oblateness model given by equation
(\ref{eq:shape1}) differs from this value only in the 
fourth significant figure. In Figure \ref{fig:L49} light curves
for light emitted from this co-latitude and observed at 
an inclination of $i=70^\circ$ are shown. The bold solid curve
labelled ``exact'' is computed using the exact numerical method
described by \cite{CMLC07}. The curve labelled ``S+D'' is 
computed for a spherical $1.4 M_\odot$ star with a radius of 15.1 km.
The curve labelled ``OS'' is computed using the Oblate Schwarzschild
approximation described in this paper. For this combination of
emission and observer angles, the S+D approximation predicts 
an eclipse for part of the rotational period. In the cases of 
the exact calculation and the OS approximation the spot is visible
all of the time. The small differences between the Exact and OS
light curves are due to differences in the spacetime metric
and to small errors due to the oblateness model given in equation
(\ref{eq:shape1}). The differences between the OS and Exact light
curves are very small compared to the differences between the
S+D and Exact light curves.

\begin{figure}
\begin{center}
\plotone{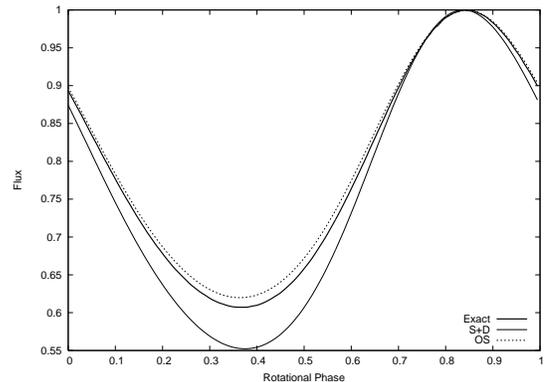}
\end{center}
\caption[]{Light curves for light emitted from a co-latitude of
  $\theta=41^\circ$ and an observer at an inclination of $i=20^\circ$.
The neutron star's parameters are exactly as given in Figure \ref{fig:L49}.
}
\label{fig:L41}
\end{figure}

In Figure \ref{fig:L41} light curves for the same star are shown, but
in this case light is emitted from a co-latitude of $\theta=41^\circ$,
and the observer's inclination is $i=20^\circ$. For this stellar
model, the radius of the star at $\theta=41^\circ$ is 14.8 km. The
light curves in Figure \ref{fig:L41} have the same meaning as in the 
previous figure, however for the S+D approximation, the surface of
the spherical star is chosen to be 14.8 km, in order to keep the
values of the gravitational potentials the same in both OS and S+D
approximations. As in the case for emission from $49^\circ$, the 
OS and Exact light curves differ by an amount which is much
smaller than the difference between the S+D and the Exact light 
curves.

\section{Conclusions}
\label{s:conclusions}

In our previous work \citep{CMLC07}  we demonstrated the need to 
include the oblateness of rapidly rotating neutron star when 
computing light curves. In this paper we presented a simple
method, the Oblate Schwarzschild approximation
which can be used to account for the star's oblate shape. 
The method presented in 
sections~\ref{s:oblate} and \ref{s:light} can be summarized as follows. First the mass, radius
and angular velocity of the neutron star are chosen. The shape 
of the star's surface is then given by 
the empirical formula given in equation (\ref{eq:shape1}),
which in turn allows the calculation of the angles
$\gamma$ and $\beta$, where $\gamma$ is the angle between 
the surface normal and the radial direction, and $\beta$
is the angle between the initial photon direction and
the surface normal. Given the angle $\beta$, the light
curve for the rapidly rotating neutron star can be 
computed using the same formulae used in the S+D
approximation (as given, for example by \cite{PG03})
with the following changes: the angle $\alpha$ (between
the initial photon direction and the radial direction)
is replaced by $\beta$; the visibility condition is
given by $0\le \cos\beta \le 1$; and the photons 
must be binned by their times of arrival. In the OS
approximation the bending angles and times of arrival
are calculated using the Schwarzschild metric. The 
resulting light curves closely approximate the 
exact light curves, as shown in section \ref{s:compare}.

The accreting ms X-ray pulsars such as \saxj
\citep{WvdK98,CM98} 
are rotating rapidly enough that 
light curve models should include the oblateness model presented in
this paper along with realistic models of the emission
spectrum and geometry. Comparisons of the light curves 
arising from these oblate models with data for \saxj
are currently in progress \citep{lea07}.

\acknowledgments
This research was supported by grants from NSERC. 
We thank Mark Alford and Ben Owen for providing us with equation
of state files.
S.~M.~M. thanks the Pacific Institute for Theoretical Physics
and the Department of Physics \& Astronomy at the University 
of British Columbia for hospitality during her sabbatical.

\begin{deluxetable}{lll}
\tablecaption{Neutron Star Shape Parameters}
\tablehead{
\colhead{Coefficient} & \colhead{Neutron and Hybrid Quark Stars}
& \colhead{CFL Quark Stars} }
\startdata
$a_0$ & $-0.18 \epsilon + 0.23 \zeta \epsilon  - 0.05 \epsilon^2$&
        $-0.26 \epsilon + 0.50 \zeta \epsilon  - 0.04 \epsilon^2$\\
$a_2$ & $-0.39 \epsilon + 0.29 \zeta  \epsilon + 0.13 \epsilon^2 $ &
        $-0.53 \epsilon + 0.85 \zeta \epsilon + 0.06 \epsilon^2$ \\
$a_4$ & $+0.04 \epsilon -0.15 \zeta \epsilon + 0.07 \epsilon^2$&
        $+0.02 \epsilon -0.14 \zeta \epsilon + 0.09 \epsilon^2$\\
$b_0$ & $+0.18 \tily -0.23 \tilx\tily + 0.18 \tily^2$&
        $+0.26 \tily -0.50 \tilx\tily + 0.31 \tily^2$\\
$b_2$ & $+0.39\tily -0.29 \tilx \tily  + 0.42 \tily^2$&
        $+0.53\tily -0.85 \tilx \tily  + 1.06 \tily^2$\\
$b_4$ & $-0.04 \tily + 0.15 \tilx \tily - 0.13 \tily^2$&
        $-0.02 \tily + 0.14 \tilx \tily - 0.12 \tily^2$\\
$c_2$ & $+0.60 \tily^2$&
        $+1.13 \tily^2$\\
$c_4$ & $-0.12 \tily^2$&
        $-0.07 \tily^2$\\
\enddata
\end{deluxetable}


\begin{thebibliography}{}
\bibitem[Alford et al.(2005)]{ABPR}
Alford, M., Braby, M., Paris, M., \& Reddy, S. 2005,
ApJ 629, 969
%
\bibitem[Akmal et al.(1998)]{APR98}
Akmal, A., Pandharipande, V. R., \& Ravenhall, D. G. 1998,
Phys. Rev. C, 58, 1804
%
\bibitem[Arnett \& Bowers(1977)]{AB77} 
Arnett, W.~D., \& Bowers, R.~L. 1977, \apjs, 33, 415
%
\bibitem[Beloborodov(2002)]{Bel02}
Beloborodov, A. M. 2002, \apj, 566, L85
%
\bibitem[Bhattacharyya et al.(2005)]{Bhat05} Bhattacharyya, 
S., Strohmayer, T.~E., Miller, M.~C., \& Markwardt, C.~B.\ 2005, \apj, 619, 
483 
%
\bibitem[Bogdanov et al.(2006)]{Bog07} Bogdanov, S., Rybicki, 
G.~B., \& Grindlay, J.~E.\ 2006, arXiv:astro-ph/0612791 
%
\bibitem[Braje \& Romani(2001)]{Braje01} Braje, T.~M., \& 
Romani, R.~W.\ 2001, \apj, 550, 392
%
\bibitem[Braje et al.(2000)]{Braje00} Braje, T.~M., Romani, 
R.~W., \& Rauch, K.~P.\ 2000, \apj, 531, 447
%
\bibitem[Cadeau et al.(2005)]{CLM05}
Cadeau, C., Leahy, D.~A., \& Morsink, S.~M. 2005, \apj, 618, 451
%
\bibitem[Cadeau et al.(2007)]{CMLC07}
Cadeau, C., Morsink, S. M., Leahy, D. A., \& Campbell, S.~S. 2007,
\apj, 654, 458
%
\bibitem[Chakrabarty \& Morgan(1998)]{CM98} Chakrabarty, D., 
\& Morgan, E.~H.\ 1998, \nat, 394, 346 
%
\bibitem[Cook, Shapiro, \& Teukolsky(1994)]{CST94}
Cook, G. B., Shapiro, S. L., Teukolsky, S. A. 1994, \apj, 424, 823
%
\bibitem[Fraga et al.(2001)]{Fraga} Fraga, E.~S., Pisarski, 
R.~D., \& Schaffner-Bielich, J.\ 2001, \prd, 63, 121702 
%
\bibitem[Glendenning(2000)]{Glen} Glendenning, N.~K.\ 2000, 
{\em Compact Stars}, (Springer-Verlag, New York).
%
\bibitem[Hartle(1967)]{Hartle67}
Hartle, J. 1967, ApJ 150, 1005
%
\bibitem[Lackey, Nayyar, \& Owen(2006)]{LNO06} 
Lackey, B.~D., Nayyar, M., \& Owen, B.~J.\ 2006, \prd, 73, 024021 
%
\bibitem[Leahy(2004)]{lea04} 
Leahy, D.~A. 2004, \apj, 613, 517
%
\bibitem[Leahy et al.(2007)]{lea07}
Leahy, D. A., Morsink, S. M., \& Cadeau, C. 2007, submitted to ApJ,
astro-ph/0703287
%
\bibitem[Miller \& Lamb(1998)]{ML98}
Miller, M.~C. \& Lamb, F.~K. 1998, \apjl, 499, L37
%
\bibitem[Misner, Thorne, \& Wheeler(1973)]{MTW}
Misner, C.~W., Thorne, 
K.~S., \& Wheeler, J.~A. 1973, Gravitation, 
W.H.~Freeman and Co.
%
\bibitem[Pechenick, Ftaclas, \& Cohen(1983)]{PFC83}
Pechenick, K.~R., Ftaclas, C., \& Cohen, J.~M. 1983, \apj, 274, 846
%
\bibitem[Poutanen \& Beloborodov(2006)]{PB06} 
Poutanen, J., \& Beloborodov, A.~M.\ 2006, \mnras, 373, 836
%
\bibitem[Poutanen \& Gierli\'{n}ski(2003)]{PG03}
Poutanen, J. \& Gierli\'{n}ski, M. 2003, \mnras, 343, 1301
%
\bibitem[Stergioulas \& Friedman(1995)]{SF95}
Stergioulas, N. \& Friedman, J.~L. 1995, \apj, 444, 306
%
\bibitem[Wijnands \& van der Klis(1998)]{WvdK98} Wijnands, R., 
\& van der Klis, M.\ 1998, \nat, 394, 344
\end{thebibliography}
\end{document}